\documentclass[10pt,twocolumn,letterpaper]{article}

\usepackage{cvpr}              

\usepackage[dvipsnames]{xcolor}

\definecolor{cvprblue}{rgb}{0.21,0.49,0.74}
\usepackage[pagebackref,breaklinks,colorlinks,citecolor=cvprblue]{hyperref}

\def\paperID{109} 
\def\confName{CVPR}
\def\confYear{2024}

\title{Training Transformer Models by Wavelet Losses Improves \\ Quantitative and Visual Performance in Single Image Super-Resolution}

\author{Cansu Korkmaz, A. Murat Tekalp\\ 
College of Engineering and KUIS AI Center, Koc University\\
{\tt https://github.com/mandalinadagi/Wavelettention}
}

\begin{document}
\maketitle
\begin{abstract}
Transformer-based models have achieved remarkable results in low-level vision tasks including image super-resolution (SR). However, early Transformer-based approaches that rely on self-attention within non-overlapping windows encounter challenges in acquiring global information. To activate more input pixels globally, hybrid attention models have been proposed. Moreover, training by solely minimizing pixel-wise RGB losses, such as $l_1$, have been found inadequate for capturing essential high-frequency details. This paper presents two contributions: i) We introduce convolutional non-local sparse attention (NLSA) blocks to extend the hybrid transformer architecture in order to further enhance its receptive field. ii) We employ wavelet losses to train Transformer models to improve quantitative and subjective performance. While wavelet losses have been explored previously, showing their power in training Transformer-based SR models is novel. Our experimental results demonstrate that the proposed model provides state-of-the-art PSNR results as well as superior visual performance across various benchmark datasets. 
\end{abstract}
\vspace{-10pt}    
\section{Introduction}
\label{sec:intro}
Single-image super-resolution (SR) aims to recover high-frequency (HF) details that are missing in low-resolution~(LR) images. Early deep-learning-based models used simple convolutional neural networks (CNN) \cite{dong_srcnn, kim2016accurate}. Subsequently, various methods introduced residual learning, such as Enhanced deep residual networks (EDSR) \cite{EDSR2017}, and attention mechanisms, such as residual channel attention networks (RCAN) \cite{RCAN2018}. More advanced models integrated dense connections \cite{tong_densenet, zhang_res_dense}, spatial and channel attention networks \cite{RCAN2018, dai_attention, niu_han, mei_nlsa, zhang2021tsan}, yielding remarkable performance in terms of peak signal-to-noise ratio (PSNR) and structural similarity measure (SSIM). However, CNNs encounter challenges in modeling long-range dependencies, prompting the development of Transformer-based image SR networks \cite{niu_han, Liang2021SwinIRIR, cat_chen2022cross, art_zhang2023accurate, chen2023activating}. For instance, SwinIR \cite{Liang2021SwinIRIR}, based on the Swin Transformer \cite{st_9710580}, significantly enhanced SR performance. Additionally, a hybrid attention transformer (HAT) \cite{chen2023activating}, which combines channel attention and window-based self-attention with an overlapping cross-attention module, achieved state-of-the-art results.

While Transformer-based SR models demonstrate impressive performance, there is still room for further improvement. For example, HAT~\cite{chen2023activating} provides improved performance over SwinIR~\cite{Liang2021SwinIRIR} by introducing an overlapping cross-attention module to activate more pixels capturing longer-range dependencies. Recent studies \cite{conv_vit_xiao2021early, conv_vit_yuan2021incorporating} have shown that integrating early convolutional layers can enhance visual representation in Transformer-based models. Based on this observation, we propose extending the~HAT~\cite{chen2023activating} architecture by incorporating non-local attention (NLSA) blocks \cite{mei_nlsa} to further expand the receptive field and enhance the quality of reconstructed SR images.

It is well-established that merely minimizing RGB domain pixel-wise $l_1$ or $l_2$ losses is inadequate for capturing high-frequency details essential for achieving visually pleasing results~\cite{wang2018esrgan, wavelet_face, wavelet_autoenc, korkmaz2024training}. Since high-frequency image details are well represented by wavelet coefficients, we employ an additional wavelet loss term during the training 
to assist reconstruction of high-frequency details and improve both PSNR and visual quality of resulting SR images.

In summary, our main contributions are:
\begin{itemize}
    \item We propose a new hybrid Transformer-based architecture for image SR that integrates convolutional non-local self-attention (NLSA) blocks with a Transformer-based model aimed at expanding the receptive field of the model. 
    \item We introduce a wavelet loss term for training that enables SR models to better capture  high-frequency image details.
    \item Training the proposed model by the wavelet loss not only improves the PSNR but also the visual quality of images. In particular, we obtain up to 0.72 dB PSNR gain over the state-of-the-art HAT model on the Urban100 test set.
    \item The proposed framework is generic in the sense that any Transformer-based SR network can be plugged into this framework and trained by wavelet losses for better results.
\end{itemize}
\section{Related Work}

\subsection{Convolutional Attention Models} \vspace{-3pt}
Several popular SR models employ convolutional attention mechanisms \cite{RCAN2018, dai_attention, niu_han, zhang2021tsan, mei_nlsa, attention_zhang2022efficient}. One of the~pioneer works, deep residual channel attention networks~(RCAN)~\cite{RCAN2018} proposed high-frequency channel-wise feature attention using a residual-in-residual (RIR) structure to form a deep CNN architecture. Second-order attention network (SAN) \cite{dai_attention}  proposed a trainable second-order channel attention module for enhanced feature learning and adaptive channel-wise feature rescaling based on second-order statistics. SAN also integrated non-local operations for long-distance spatial context and features for learning progressively abstract feature representations. Holistic attention network (HAN) \cite{niu_han} proposed capturing the correlation among different convolution layers with layer attention and channel-spatial attention modules. The Non-Local Sparse Attention (NLSA)~\cite{mei_nlsa} method introduced a dynamic sparse attention pattern, combining long-range modeling from non-local operations with robustness and efficiency of sparse representation. Recently proposed Efficient Long-Range Attention Network (ELAN) \cite{attention_zhang2022efficient} incorporated a shift convolution (shift-conv) method and group-wise multi-scale self-attention module to preserve local structural information and long-range image dependencies. In this work, we incorporated NLSA blocks into a window-based transformer SR model for enlarged receptive field.

\subsection{Transformer-Based SR Models} \vspace{-3pt}
Transformer models, which are extremely successful  in natural language processing~\cite{NIPS2017_attention_all_u_need}, have also been adopted by the~computer vision community for high-level vision tasks~\cite{vit_chu2021Twins, vit_chu2023CPVT, vit_Dong_2022_CVPR, vit_li2021localvit, vit_pyramid, vit_wu2021cvt, vit_wu2022pale, liu2024pasta}, as well as low-level vision tasks including image SR \cite{vit_ll_chen2021ipt, vit_ll_cao2021video, vit_ll_tian2024cross, vit_ll_wang2022uformer, vit_ll_chen2023learning, Liang2021SwinIRIR, LSDIR_Li_2023_CVPR, cat_chen2022cross, art_zhang2023accurate, chen2023activating, li2023feature}. 

Among the popular transformer-based image SR models, IPT \cite{vit_ll_chen2021ipt} introduced a vision Transformer style network with multi-task pre-training for image SR. SwinIR \cite{Liang2021SwinIRIR} proposed an image restoration method based on the Swin Transformer \cite{st_9710580}. UFormer \cite{vit_ll_wang2022uformer} employed a locally-enhanced Transformer block to reduce computational complexity and introduced a learnable multi-scale restoration modulator for capturing better both local and global dependencies for image restoration. CAT \cite{cat_chen2022cross} developed rectangle-window self-attention and a locality complementary module to enhance image restoration; later, CRAFT \cite{li2023feature} proposed the high-frequency enhancement residual block along with a fusion strategy for Transformer-based SR methods to improve performance. ART \cite{art_zhang2023accurate} enlarged the receptive field using an attention retractable module for improved SR performance. SRFormer \cite{transformer_Zhou_2023_ICCV} suggested permuted self-attention that effectively balances channel and spatial information, enhancing the performance of self-attention mechanisms. Dual Attention Transformer (DAT) \cite{transformer_Chen_2023_ICCV} combined spatial and channel feature aggregation both between and within Transformer blocks, enabling global context capture and inter-block feature aggregation. More recently, HAT~\cite{chen2023activating} combined channel attention and window-based self-attention for improved global and local feature utilization, achieving state-of-the-art results in image SR tasks. However, existing Transformer-based models still do not fully harness global image correlations and there is room for further improvements \cite{ntire23_zhu2023attention}. To this effect, we propose to leverage more input pixels by cascading convolutional attention models with window-based transformer SR models aiming to enhance image reconstruction quality.

\subsection{Frequency/Wavelet-Domain Models/Losses} \vspace{-3pt}
Several researchers have recognized the need to treat low and high frequency image features differently and delved into frequency/wavelet domain methodologies to tackle image restoration/SR tasks \cite{freq_pang2020fan, xin2020wavelet, freq_zhang2022swinfir, sheng2022frequency, fu2021dw}. Some of them explored the decomposition of features into frequency bands using multi-branch CNN architectures in the Fourier domain~\cite{freq_baek2020single}. Frequency aggregation networks~\cite{freq_pang2020fan} extracted various frequencies from LR images and feed them into a channel attention-grouped residual dense network to recover HR images with enhanced details and textures. Other approaches include use of Fourier-domain architectures~\cite{Fuoli2021FourierSL, dualformer_luo2023effectiveness, freq_zhang2022swinfir} and Fourier-domain loss functions~\cite{jiang2021focal}. For instance, SwinFIR~\cite{freq_zhang2022swinfir} extended~SwinIR by incorporating image-wide receptive fields using fast Fourier convolution, while DualFormer \cite{dualformer_luo2023effectiveness} leveraged spatial and spectral discriminators simultaneously in the Fourier domain. 
However, approaches that rely on Fourier domain methods and losses lack the ability to localize and capture the~scale/orientation of high-frequency image features, thereby show limited improvements in SR performance. 

Alternatively, wavelet-domain SR methods, such~as DWSR~\cite{DWSR_guo2017deep}, Wavelet- SRNet~\cite{wavelet_srnet_huang2017wavelet},
WDST \cite{Deng2019WaveletDS}, WIDN~\cite{WIDN_sahito2019wavelet}, 
wavelet-based dual recursive network~\cite{xin2020wavelet}, WRAN~\cite{WRAN_xue2020wavelet} and PDASR~\cite{PDASR}, predict wavelet coefficients of SR images. 
In particular, PDASR~\cite{PDASR} modified the~model architecture to condition reconstructed images on low-level wavelet subbands to reduce visible artifacts and improve performance.
In contrast, most recently, WGSR~\cite{korkmaz2024training} proposed training  RGB-domain standard GAN-SR models using a weighted combination of wavelet subband losses, departing from conventional RGB $l_1$ loss to control artifacts. 
In this paper, we show that training hybrid transformer SR models by wavelet losses also improves their performance. Our results are superior to others because predicting RGB pixels is easier than predicting sparse wavelet coefficients of detail bands, while unequal weighting of losses in different wavelet subbands enables learning structures with different scales and orientations.
\begin{figure*}
\centering
\includegraphics[width=\linewidth]{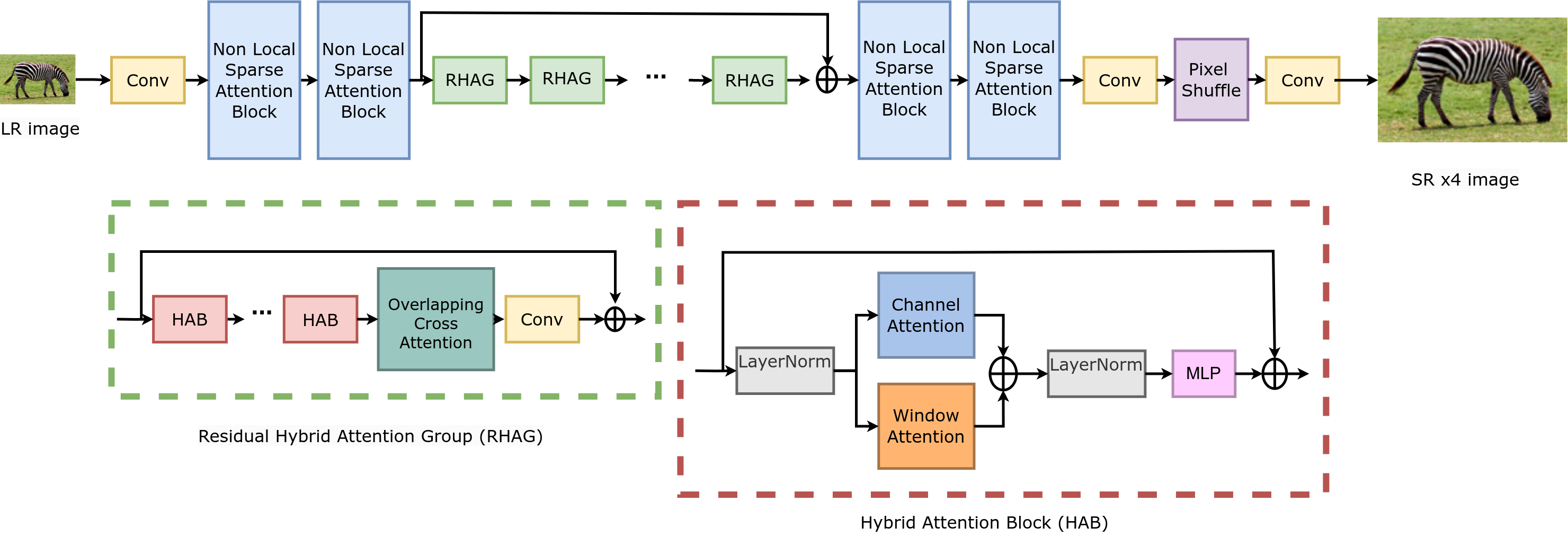}  \vspace{-16pt}
\caption{The proposed image SR architecture, which sandwiches HAT \cite{chen2023activating} in between NLSA blocks \cite{mei_nlsa} for enlarged receptive field.}
\label{fig:arch}
\end{figure*}

\section{Proposed Method}
\subsection{Hybrid Attention Architecture}
The proposed approach enhances the HAT architecture~\cite{chen2023activating} by sandwiching it in between non-local sparse attention (NLSA) blocks \cite{mei_nlsa} as depicted in Figure \ref{fig:arch}. The baseline HAT architecture utilized residual-in-residual approach \cite{wang2018esrgan} and developed hybrid attention block (HAB) similar to the standard Swin Transformer block \cite{Liang2021SwinIRIR} to enhance performance. In addition, the NLSA module \cite{mei_nlsa} combines several techniques to enhance efficiency and global modeling in attention mechanisms. Specifically, NLSA uses Spherical Locality Sensitive Hashing method to divide input features into buckets to calculate attention. This module incorporates the strengths of Non-Local Attention, which allows for global modeling and capturing long-range dependencies in the image. Additionally, it leverages advantages of sparsity and hashing \cite{mei_nlsa}, which lead to high computational efficiency. Therefore, by sandwiching the HAT \cite{chen2023activating} architecture in between 2 or 4 NLSA blocks, we aim to increase the effectiveness of the attention mechanism, making the framework suitable for various applications in deep learning models especially for SR.

\subsection{Training Loss Function}

\begin{figure}
\centering
\includegraphics[width=\linewidth]{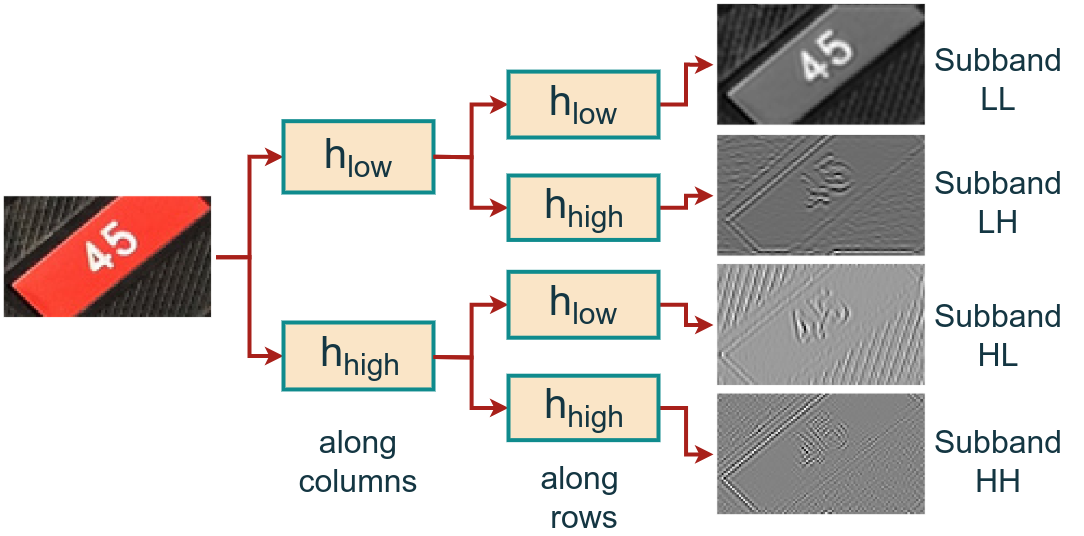} 
\caption{Illustration of the Stationary Wavelet Transform (SWT). SWT uses low-pass and high-pass decomposition filter pairs to compute the wavelet coefficients without subsampling subbands.}
\label{fig:swt}
\end{figure}

The wavelet loss is proposed to capture high-frequency details essential for visually pleasing SR results. This modified hybrid Transformer-based SR architecture is trained using wavelet losses along with the conventional $l_1$ RGB loss. The Stationary Wavelet Transform (SWT) is a technique that enables the multi-scale decomposition of images~\cite{wavelet_doc}. The illustration of SWT decomposition is depicted in Figure \ref{fig:swt}, which results in one low-frequency (LF) subband, called LL, and multiple high-frequency (HF) subbands, called LH, HL, and HH. The number of HF subbands is determined by the number of decomposition levels, and each HF subband contains detailed information in one of horizontal, vertical, or diagonal directions. SWT inherently combines scale/frequency information with spatial location, making it particularly suitable for tasks where preserving spatial details across different scales is essential, such as in SR applications. This combination of SWT loss with the proposed Transformer model aims to improve the overall performance of the baseline HAT model \cite{chen2023activating} by incorporating non-local attention mechanisms and leveraging wavelet losses to enhance image quality during training. To the best of our knowledge, this is the first work to use a wavelet loss function to train Transformer-based image SR models.

\begin{figure}
\centering
\includegraphics[width=\linewidth]{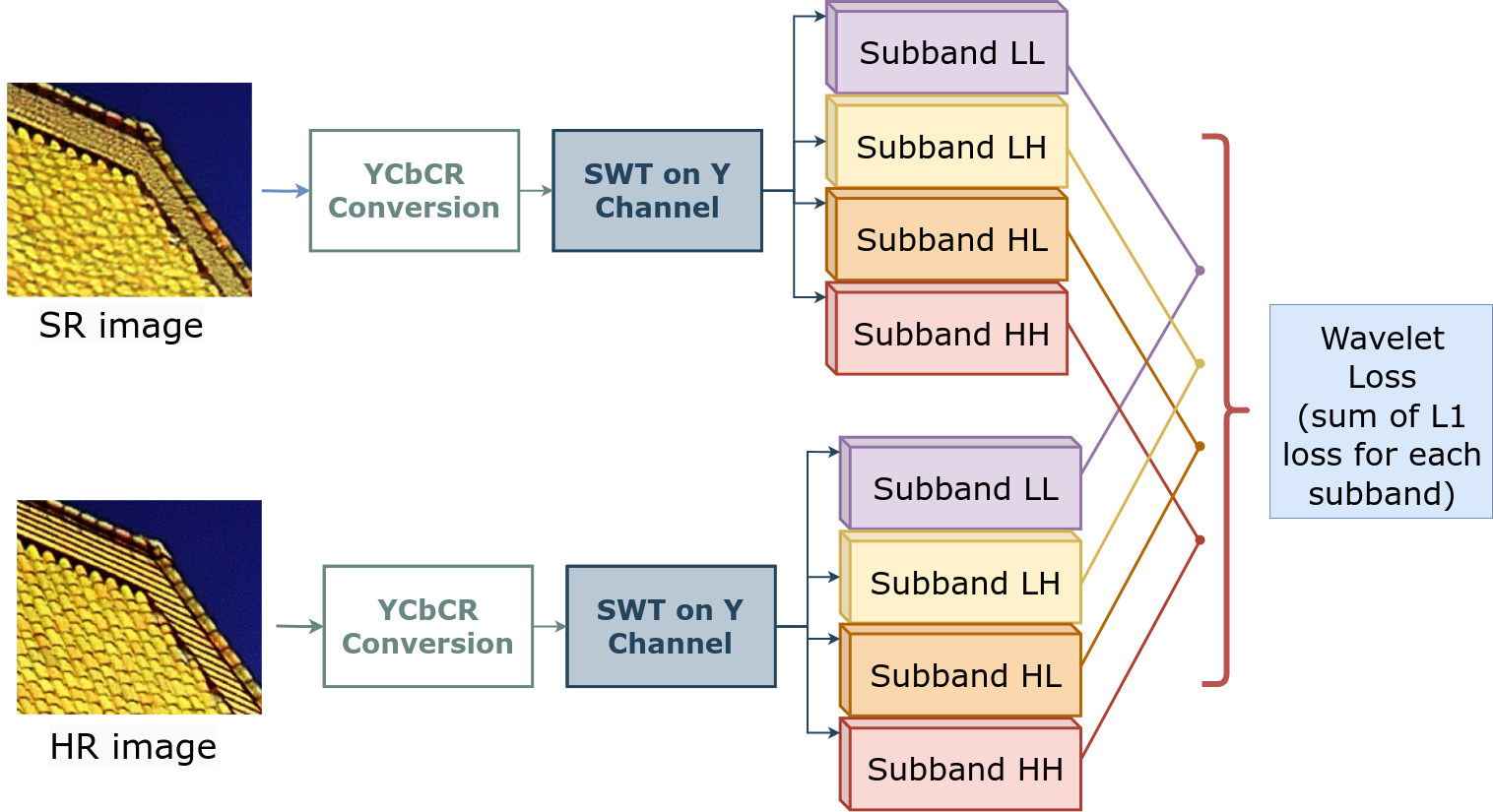} 
\caption{Computation of pixel-wise $l_1$ loss on SWT subbands. Training hybrid transformer architectures by a weighted combination of RGB and SWT losses results in remarkable quantitative and qualitative performance improvements.}
\label{fig:loss}
\end{figure} 

SWT is known for its efficiency and intuitive ability to represent and store multi-resolution images effectively \cite{wavelet_96, wavelet_doc}. This means that it can capture both contextual and textural information of an image across different levels of detail. The understanding of how WT operates and its capacity to handle varying levels of image details inspired us to integrate wavelet losses into a Transformer-based super-resolution system. In other words, wavelet subbands have capability of handling different aspects of information encoded in images that enables it a promising addition to enhance the performance of generated SR images. Hence, rather than employing the typical RGB-domain fidelity loss seen in conventional Transformer methods, we introduce the SWT fidelity loss denoted as $L_{SWT}$ along with corresponding tuning parameter. The $l_1$ fidelity loss in wavelet-domain is calculated between the SWT subbands of the generated images $x$ and the ground truth (GT) image $y$, is illustrated in Figure \ref{fig:loss}. The total wavelet loss is averaged over a minibatch size represented by $\mathbb{E}[.]$. This formulation allows for a more nuanced optimization process that can enhance the overall quality of the super-resolved images produced by the image Transformer models.

\small
\begin{multline}  \label{eq:swt_fidelity}
 L_{SWT} = \mathbb{E} \bigl[ \sum_{j} \lambda_{j} \big\| {SWT(G(x))}_{j} - {SWT(y)}_{j} \big\| _1  \bigr]   
\end{multline} 
\normalsize 
where $G$ denotes the proposed Transformer-based SR model and $\lambda_j$ are appropriate scaling factors to control the generated HF details. 

\noindent
The overall loss for the training is given by 
\begin{equation}
 L_{G} = L_{RGB} + L_{SWT}
\end{equation} 
where $L_{RGB}$ denotes the $l_1$ fidelity loss calculated on RGB domain, measuring pixel-wise errors in the image space. 

The wavelet $l_1$ loss landscape differs from Y channel $l_1$ loss if i) we use a non-orthogonal wavelet transform, and/or ii) we use different weights for different subbands. It is worth noting that failing to adjust the balance between the wavelet subband loss terms and the RGB fidelity loss term (using scaling parameters $\lambda_j$) may result in chroma artifacts (producing greenish images), as the wavelet loss is computed based on the Y-channel only and favors Y channel. An alternative approach to address this potential issue would be incorporating an additional chroma loss term to explicitly preserve the color balance.

\section{Experimental Results}
\subsection{Implementation Details}
\label{subsec:implementation_details}

We configured the chunk size for the non-local sparse attention as 144 and added 2 consecutive NLSA \cite{mei_nlsa} layers before and after the HAT \cite{chen2023activating} architecture. We utilized pre-trained HAT-L with default configuration. Hence the embedding dimension is set to 180 and patch embedding is set to 4. The total parameter number of proposed method is 41.3M. During training, we randomly crop 64x64 patches from the LR images from LSDIR \cite{LSDIR_Li_2023_CVPR} and DIV2K \cite{Agustsson_2017_CVPR_Workshops} datasets to form a mini-batch of 8 images. The training images are further augmented via horizontal flipping and random rotation of 90, 180, and 270 degrees. We optimize the model by ADAM optimizer \cite{Kingma2014AdamAM} with default parameters. The learning rate is set to $4e^5$ and reduced by half after 125k, 200k and 240k iterations. The final model is obtained after 250k iterations. While calculating the pixel-wise loss of wavelet subbands, we utilize a Symlet filter ``sym19" to compute wavelet coefficients. In the computation of the wavelet loss (Eqns. 1 and 2), we set all $\lambda_j=0.05$ in order to avoid chroma artifacts.
Our model is implemented with PyTorch and trained on Nvdia A40 GPUs. 

\begin{table*}
 \caption{Quantitative comparison of the proposed wavelet decomposition-based optimization objective vs. other state-of-the-art methods for $\times$4 SR task. The best and the second-best are marked in \textbf{bold} and \underline{underlined}, respectively.}
    \centering
    \scalebox{0.9}{
    \begin{tabular}{lcccccccc}
    \specialrule{.1em}{.05em}{.05em} 
    Benchmark & \hspace{40pt} Set5 & & \hspace{40pt} Set14 & & \hspace{40pt} BSD100 & & \hspace{40pt}  Urban100 & \\  
    \hline
    Method & PSNR & SSIM & PSNR & SSIM & PSNR & SSIM & PSNR & SSIM \\ \hline\hline
    EDSR & 32.46 & 0.8968 & 28.80 & 0.7876 & 27.71 & 0.7420 & 26.64 & 0.8033 \\
    RCAN & 32.63 & 0.9002 & 28.87 & 0.7889 & 27.77 & 0.7436 & 26.82 & 0.8087 \\
    SAN & 32.64 & 0.9003 & 28.92 & 0.7888 & 27.78 & 0.7436 & 26.79 & 0.8068 \\
    HAN & 32.64 & 0.9002 & 28.90 & 0.7890 & 27.80 & 0.7442 & 26.85 & 0.8094 \\
    NLSA & 32.59 & 0.9000 & 28.87 & 0.7891 & 27.78 & 0.7444 & 26.96 & 0.8109 \\ 
    ELAN & 32.75 & 0.9022 & 28.96 & 0.7914 & 27.83 & 0.7459 & 27.13 & 0.8167 \\
    \hline
    SwinIR & 32.92 & 0.9044 & 29.09 & 0.7950 & 27.92 & 0.7489 & 27.45 & 0.8254 \\
    CAT-R & 32.89 & 0.9044 & 29.13 & 0.7955 & 27.95 & 0.7500 & 27.62 & 0.8292 \\
    CAT-A & 33.08 & 0.9052 & 29.18 & 0.7960 & 27.99 & 0.7510 & 27.89 & 0.8339 \\
    CRAFT & 32.52 & 0.8989 &  28.85 & 0.7872 & 27.72 & 0.7418 & 26.56 & 0.7995 \\
    ART & 33.04 & 0.9051 & 29.16 & 0.7958 & 27.97 & 0.7510 & 27.77 & 0.8321 \\
    SRFormer & 32.93 & 0.9041 & 29.08 & 0.7953 & 27.94 &  0.7502 & 27.68 & 0.8311 \\
    SRFormer+ & 33.09 & 0.9053 & 29.19 & 0.7965 & 28.00 & 0.7511 & 27.85 & 0.8338 \\
    DAT & 33.08 &  0.9055 & 29.23 &  0.7973 & 28.00 & 0.7515 &  27.87 &  0.8343 \\
    DAT+ & \underline{33.15} & \underline{0.9062} &  \underline{29.29} &  \underline{0.7983} &  \underline{28.03} & \underline{0.7518} & \underline{27.99} &  0.8365 \\ 
    HAT & 33.04 & 0.9056 & 29.22 & 0.7973 & 28.00 & 0.7517 & 27.97 & \underline{0.8368} \\
    Ours & \textbf{33.27} & \textbf{0.9082} & \textbf{29.53} & \textbf{0.8020} &\textbf{28.12} & \textbf{0.7549} & \textbf{28.69} & \textbf{0.8506} \\    
\specialrule{.1em}{.05em}{.05em} 
    \end{tabular} 
}
\label{table:quantitative_results}
\end{table*}

\subsection{Quantitative Results}

To evaluate the generalization capability of our approach, we present results on validation benchmarks such as Set5~\cite{set5_cite}, Set14 \cite{set14_cite}, BSD100 \cite{bsd100_cite}, and Urban100 \cite{urban100_cite}. We employ fidelity metrics like PSNR and SSIM to gauge the performance of our proposed method alongside various state-of-the-art image SR techniques, including traditional attention models like EDSR \cite{EDSR2017}, RCAN \cite{RCAN2018}, SAN \cite{dai_attention}, HAN \cite{niu_han}, NLSA \cite{mei_nlsa}, and ELAN \cite{attention_zhang2022efficient}. Furthermore, we conduct comparisons with the state-of-the-art Transformer-based SR methods such as SwinIR \cite{Liang2021SwinIRIR}, CAT \cite{cat_chen2022cross}, CRAFT \cite{li2023feature}, ART \cite{art_zhang2023accurate}, SRFormer \cite{transformer_Zhou_2023_ICCV}, DAT \cite{transformer_Chen_2023_ICCV}, and HAT \cite{chen2023activating}. Table \ref{table:quantitative_results} demonstrates the PSNR and SSIM performances of those methods for $\times$4 image SR task on benchmark datasets. As depicted in Table \ref{table:quantitative_results}, our method, Wavelettention, exhibits superior performance across all four benchmarks. In contrast to existing state-of-the-art Transformer-based methods such as SwinIR \cite{Liang2021SwinIRIR}, CAT \cite{cat_chen2022cross}, ART \cite{art_zhang2023accurate}, DAT \cite{transformer_Chen_2023_ICCV} and HAT \cite{chen2023activating}, our method achieves notable performance enhancements for $\times$4 SR. Particularly, our method exhibits a PSNR gain of 0.3 dB on Set14\cite{set14_cite}, 0.12 dB on BSD100 \cite{bsd100_cite}, and 0.72 dB on Urban100 \cite{urban100_cite} compared to the competitive HAT \cite{chen2023activating} method. 

{\it To summarize, the observed PSNR improvements stem from enlarged receptive field of the proposed hybrid transformer model, our training strategy including wavelet losses, and utilization of larger datasets for training. These findings affirm that our Wavelettention stands as a robust hybrid transformer-based image SR model.}

\subsection{Qualitative Results}
\begin{figure*}[!t]
\centering
\begin{subfigure}{0.1385\textwidth} 
    \begin{subfigure}{\textwidth}
        \includegraphics[width=\textwidth]{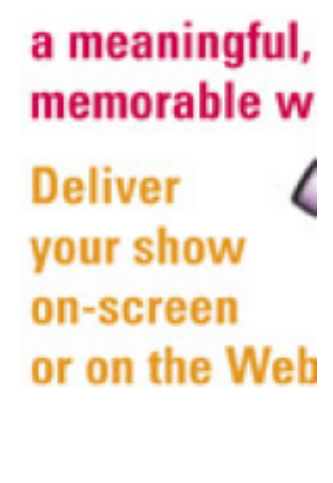} \caption*{HR - Set14}
    \end{subfigure} 
    \begin{subfigure}{\textwidth}
        \includegraphics[width=\textwidth]{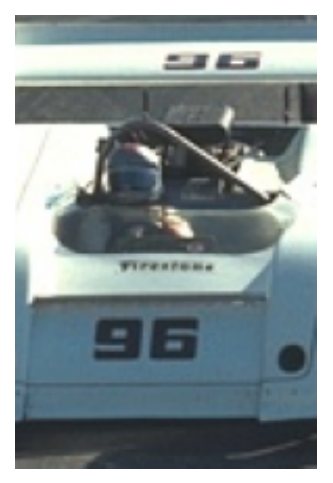}
        \caption*{HR - BSD100}
    \end{subfigure} 
    \begin{subfigure}{\textwidth}
        \includegraphics[width=\textwidth]{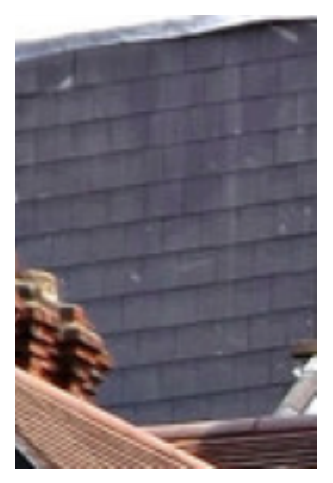}
        \caption*{HR - Urban100}
    \end{subfigure} 
    \begin{subfigure}{\textwidth}
        \includegraphics[width=\textwidth]{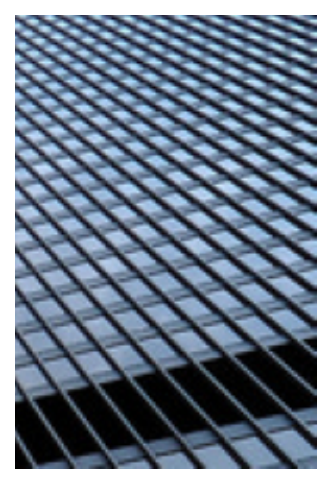}
        \caption*{HR - Urban100}
    \end{subfigure} 
    \begin{subfigure}{\textwidth}
        \includegraphics[width=\textwidth]{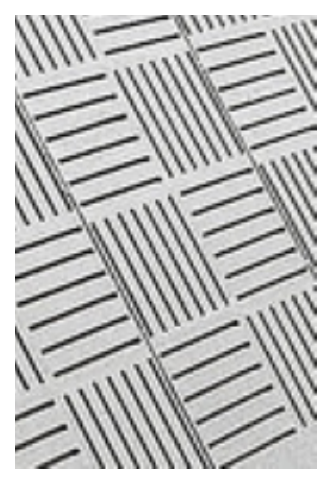}
        \caption*{HR - Urban100}
    \end{subfigure} 
\end{subfigure}
\begin{subfigure}{0.1385\textwidth} 
    \begin{subfigure}{\textwidth}
        \includegraphics[width=\textwidth]{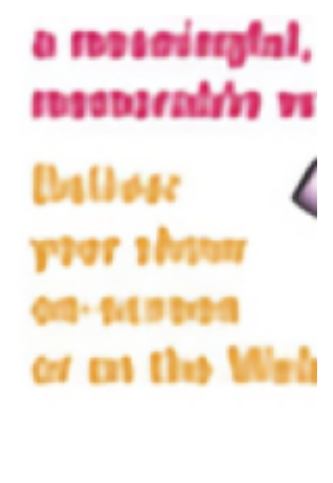}
        \caption*{SwinIR \cite{Liang2021SwinIRIR}} 
    \end{subfigure}
    \begin{subfigure}{\textwidth}
        \includegraphics[width=\textwidth]{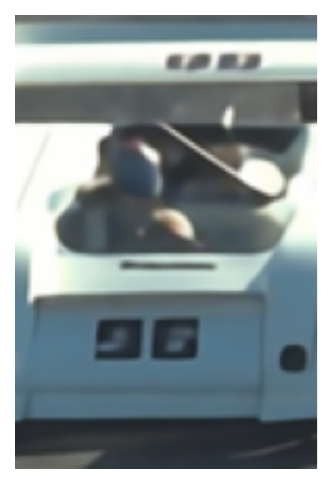}
        \caption*{SwinIR \cite{Liang2021SwinIRIR}} 
    \end{subfigure} 
    \begin{subfigure}{\textwidth}
        \includegraphics[width=\textwidth]{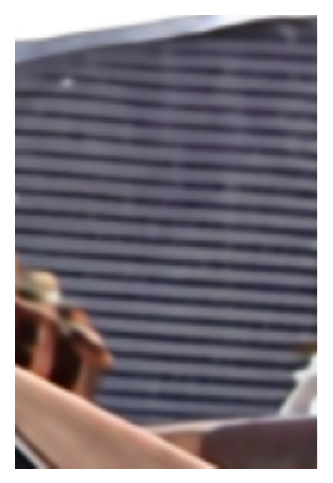}
        \caption*{SwinIR \cite{Liang2021SwinIRIR}} 
    \end{subfigure} 
    \begin{subfigure}{\textwidth}
        \includegraphics[width=\textwidth]{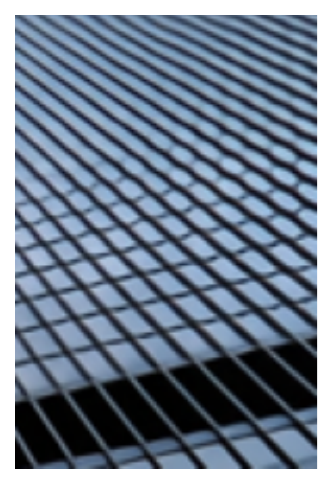}
        \caption*{SwinIR \cite{Liang2021SwinIRIR}} 
    \end{subfigure} 
    \begin{subfigure}{\textwidth}
        \includegraphics[width=\textwidth]{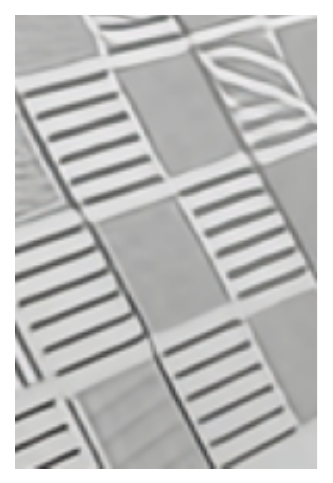}
        \caption*{SwinIR \cite{Liang2021SwinIRIR}} 
    \end{subfigure} 
\end{subfigure}
\begin{subfigure}{0.1385\textwidth} 
    \begin{subfigure}{\textwidth}
        \includegraphics[width=\textwidth]{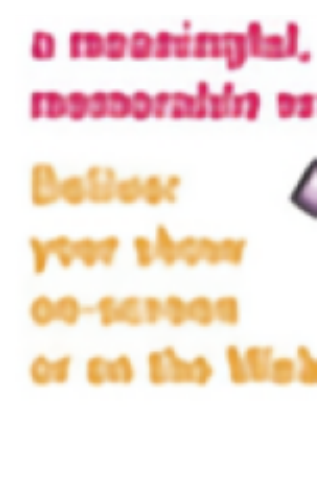}
        \caption*{CAT \cite{cat_chen2022cross}}
    \end{subfigure} 
    \begin{subfigure}{\textwidth}
        \includegraphics[width=\textwidth]{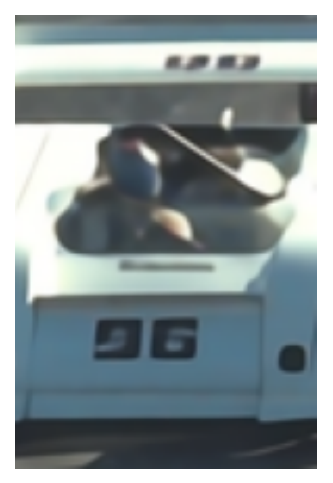}
        \caption*{CAT \cite{cat_chen2022cross}}
    \end{subfigure} 
    \begin{subfigure}{\textwidth}
        \includegraphics[width=\textwidth]{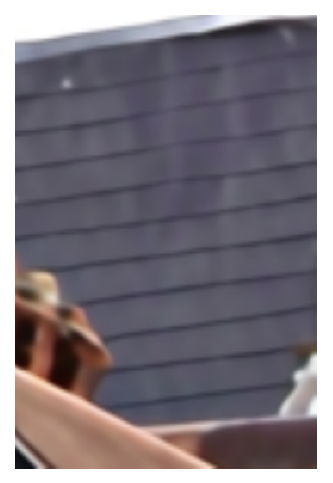}
        \caption*{CAT \cite{cat_chen2022cross}}
    \end{subfigure} 
    \begin{subfigure}{\textwidth}
        \includegraphics[width=\textwidth]{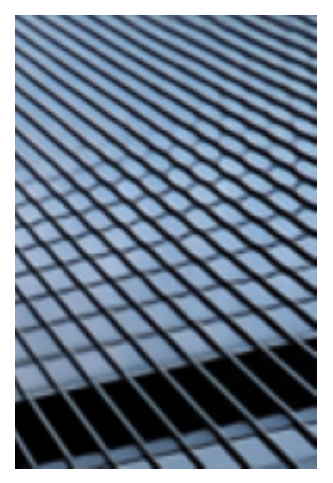}
        \caption*{CAT \cite{cat_chen2022cross}}
    \end{subfigure} 
    \begin{subfigure}{\textwidth}
        \includegraphics[width=\textwidth]{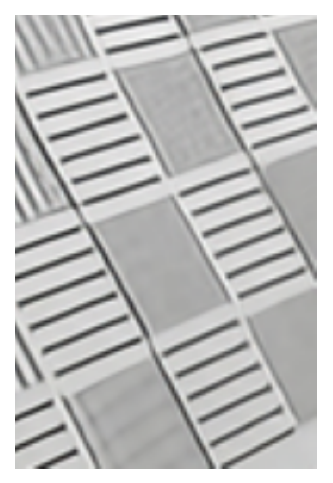}
        \caption*{CAT \cite{cat_chen2022cross}}
    \end{subfigure} 
\end{subfigure}
\begin{subfigure}{0.1385\textwidth} 
    \begin{subfigure}{\textwidth}
        \includegraphics[width=\textwidth]{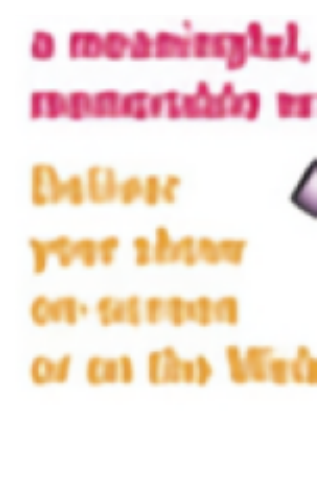}
        \caption*{ART \cite{art_zhang2023accurate}}
    \end{subfigure} 
    \begin{subfigure}{\textwidth}
        \includegraphics[width=\textwidth]{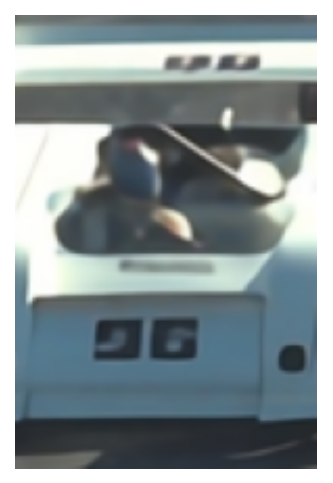}
        \caption*{ART \cite{art_zhang2023accurate}}
    \end{subfigure} 
    \begin{subfigure}{\textwidth}
        \includegraphics[width=\textwidth]{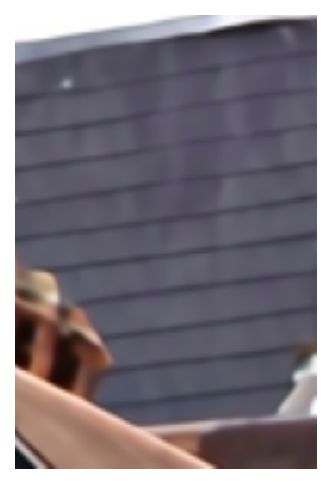}
        \caption*{ART \cite{art_zhang2023accurate}}
    \end{subfigure} 
    \begin{subfigure}{\textwidth}
        \includegraphics[width=\textwidth]{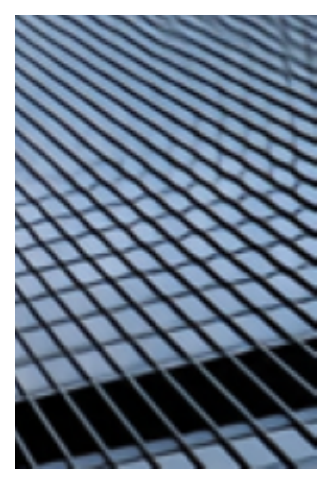}
        \caption*{ART \cite{art_zhang2023accurate}}
    \end{subfigure} 
    \begin{subfigure}{\textwidth}
        \includegraphics[width=\textwidth]{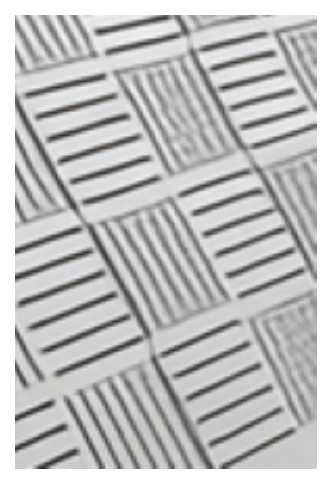}
        \caption*{ART \cite{art_zhang2023accurate}}
    \end{subfigure} 
\end{subfigure}
\begin{subfigure}{0.1385\textwidth} 
    \begin{subfigure}{\textwidth}
        \includegraphics[width=\textwidth]{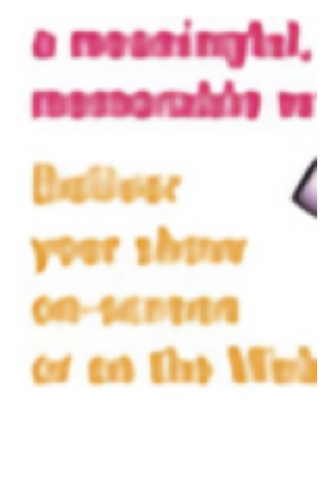}
        \caption*{DAT \cite{transformer_Chen_2023_ICCV}}
    \end{subfigure} 
    \begin{subfigure}{\textwidth}
        \includegraphics[width=\textwidth]{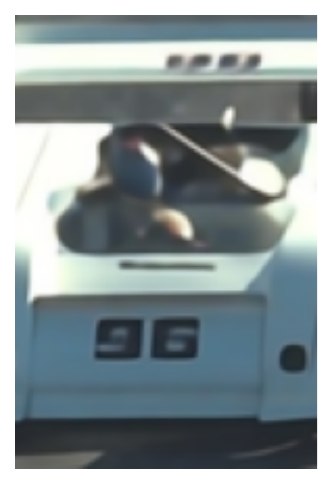}
        \caption*{DAT \cite{transformer_Chen_2023_ICCV}}
    \end{subfigure} 
    \begin{subfigure}{\textwidth}
        \includegraphics[width=\textwidth]{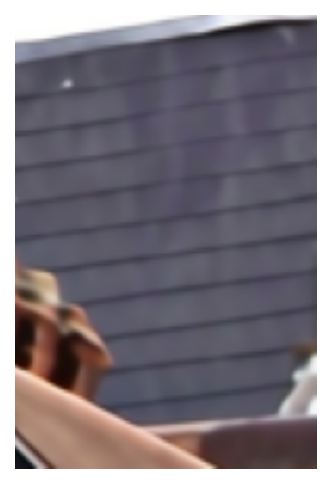}
        \caption*{DAT \cite{transformer_Chen_2023_ICCV}}
    \end{subfigure} 
    \begin{subfigure}{\textwidth}
        \includegraphics[width=\textwidth]{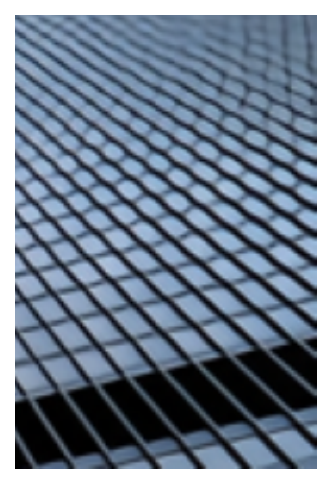}
        \caption*{DAT \cite{transformer_Chen_2023_ICCV}}
    \end{subfigure} 
    \begin{subfigure}{\textwidth}
        \includegraphics[width=\textwidth]{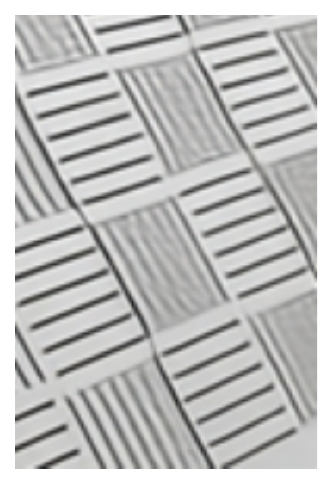}
        \caption*{DAT \cite{transformer_Chen_2023_ICCV}}
    \end{subfigure} 
\end{subfigure}
\begin{subfigure}{0.1385\textwidth} 
    \begin{subfigure}{\textwidth}
        \includegraphics[width=\textwidth]{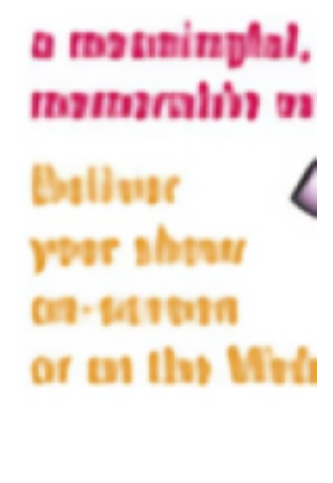}
        \caption*{HAT \cite{chen2023activating}}
    \end{subfigure} 
    \begin{subfigure}{\textwidth}
        \includegraphics[width=\textwidth]{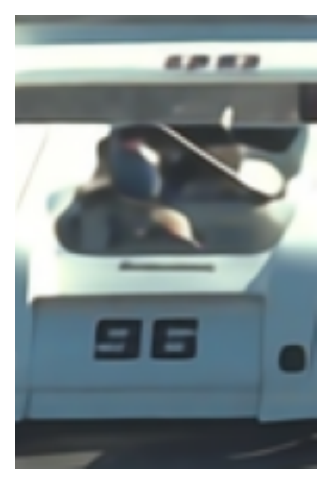}
        \caption*{HAT \cite{chen2023activating}}
    \end{subfigure} 
    \begin{subfigure}{\textwidth}
        \includegraphics[width=\textwidth]{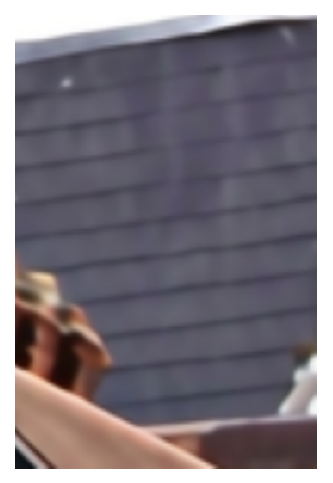}
        \caption*{HAT \cite{chen2023activating}}
    \end{subfigure} 
    \begin{subfigure}{\textwidth}
        \includegraphics[width=\textwidth]{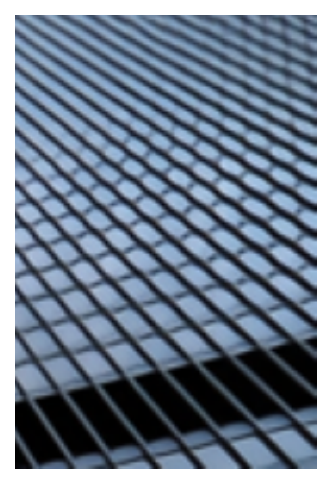}
        \caption*{HAT \cite{chen2023activating}}
    \end{subfigure} 
    \begin{subfigure}{\textwidth}
        \includegraphics[width=\textwidth]{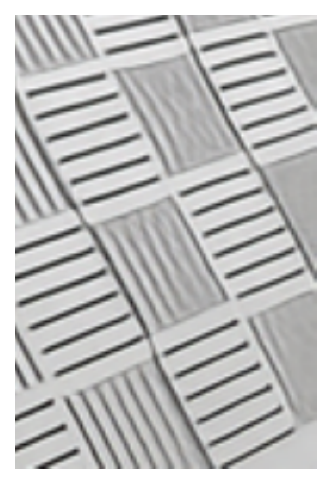}
        \caption*{HAT \cite{chen2023activating}}
    \end{subfigure} 
\end{subfigure}
\begin{subfigure}{0.1385\textwidth} 
    \begin{subfigure}{\textwidth}
        \includegraphics[width=\textwidth]{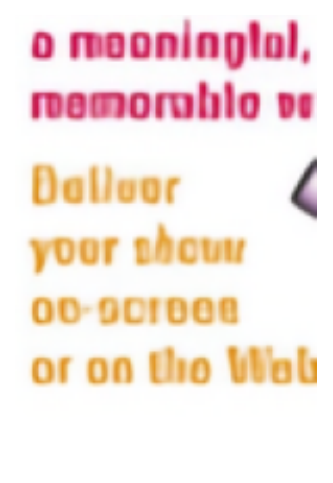}
        \caption*{Ours}
    \end{subfigure} 
    \begin{subfigure}{\textwidth}
        \includegraphics[width=\textwidth]{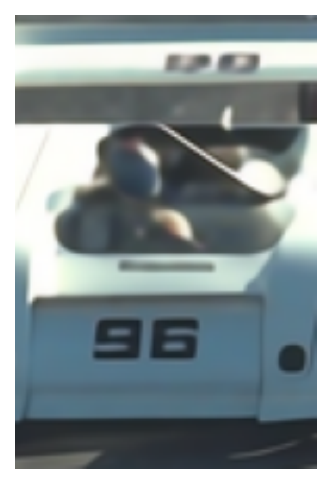}
        \caption*{Ours}
    \end{subfigure} 
    \begin{subfigure}{\textwidth}
        \includegraphics[width=\textwidth]{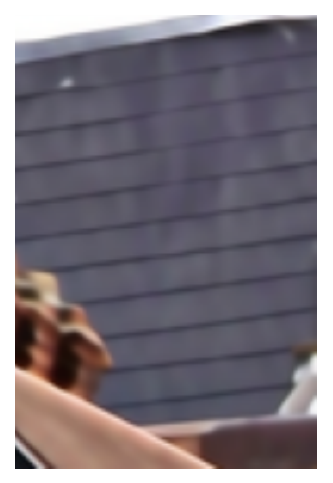}
        \caption*{Ours}
    \end{subfigure} 
    \begin{subfigure}{\textwidth}
        \includegraphics[width=\textwidth]{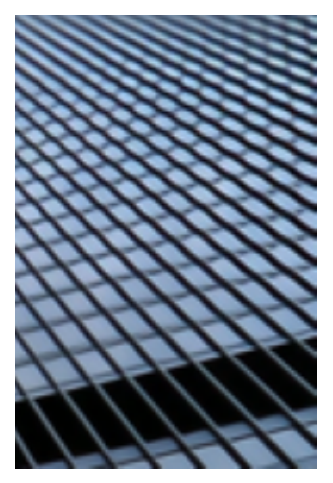}
        \caption*{Ours}
    \end{subfigure} 
    \begin{subfigure}{\textwidth}
        \includegraphics[width=\textwidth]{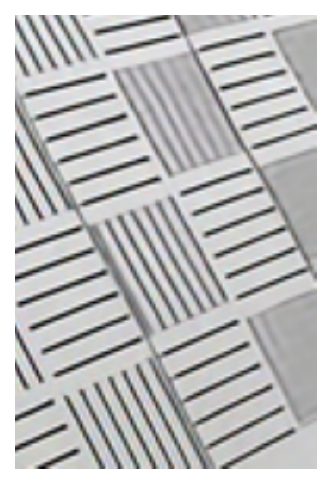}
        \caption*{Ours}
    \end{subfigure} 
\end{subfigure}
\caption{Visual comparison of $\times4$ SR results on images selected from Set14 \cite{set14_cite},~~BSD100 \cite{bsd100_cite} and~~Urban100 \cite{urban100_cite} benchmarks. Observe~that SwinIR trained by $l_1$ loss only (second column) generates aliasing artifacts, while CAT shows extreme blurring on some patches of the image in the last row. Other models show moderate blurring, while our models shows the best visual results on all images. }
\label{fig:qual_fig} 
\end{figure*}

We present challenging examples for visual comparison across three benchmark datasets in Figure \ref{fig:qual_fig}.  When compared with prominent Transformer-based methods such as SwinIR \cite{Liang2021SwinIRIR}, CAT \cite{cat_chen2022cross},
ART \cite{art_zhang2023accurate}, DAT \cite{transformer_Chen_2023_ICCV} and HAT \cite{chen2023activating}, our Wavelettention demonstrates superior restoration of detailed edges and textures. Specifically, our method exhibits a stronger ability to restore blurred text characters in Set14 \cite{set14_cite} and BSD100 \cite{bsd100_cite}. Additionally, Wavelettention successfully restores bricks of the architectures in the image patch from BSD100 \cite{bsd100_cite} which SwinIR \cite{Liang2021SwinIRIR} struggles with. Furthermore, our method manages to reconstruct the parallel stripes with small intervals in the Urban100 dataset \cite{urban100_cite}, however the other Transformer-based methods including SwinIR \cite{Liang2021SwinIRIR}, CAT \cite{cat_chen2022cross}, ART \cite{art_zhang2023accurate}, DAT \cite{transformer_Chen_2023_ICCV} and HAT \cite{chen2023activating} focus only on simpler textures due to their limited receptive field, resulting in undesirable visual outcomes. 

{\it To summarize, by sandwiching the HAT architecture in between NLSA blocks and introducing the wavelet loss, our Wavelettention model excels in the image SR task, showcasing outstanding visual performance.}

\begin{figure}[t!]
\centering
\begin{subfigure}{0.155\textwidth} 
    \begin{subfigure}{\textwidth}
        \includegraphics[width=\textwidth]{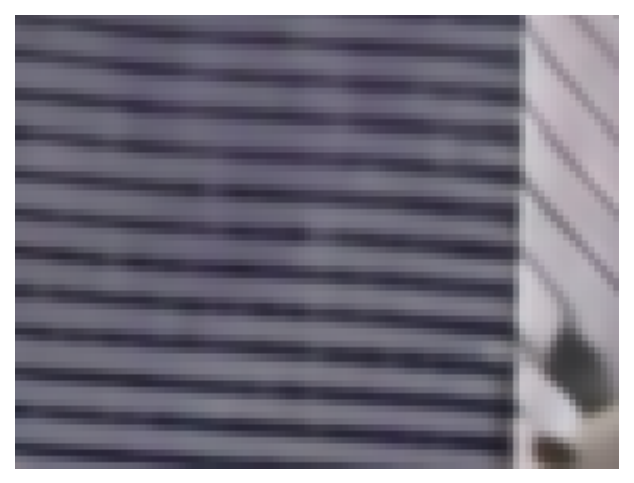} \caption*{SwinIR \cite{Liang2021SwinIRIR}}
    \end{subfigure} 
\end{subfigure}
\begin{subfigure}{0.155\textwidth} 
    \begin{subfigure}{\textwidth}
        \includegraphics[width=\textwidth]{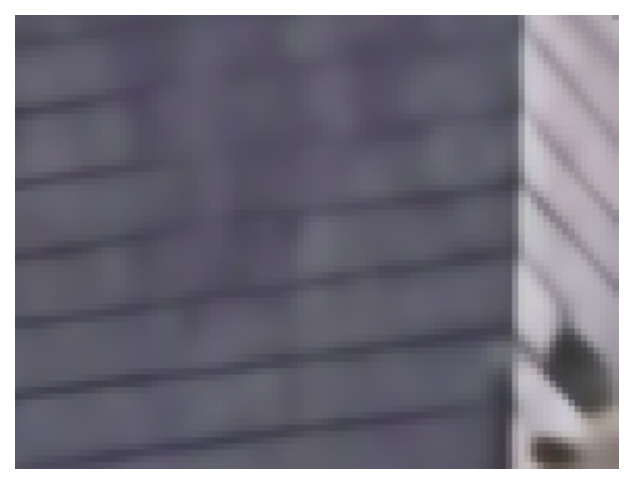}
        \caption*{SwinIR + SWT} 
    \end{subfigure}
\end{subfigure}
\begin{subfigure}{0.155\textwidth} 
    \begin{subfigure}{\textwidth}
        \includegraphics[width=\textwidth]{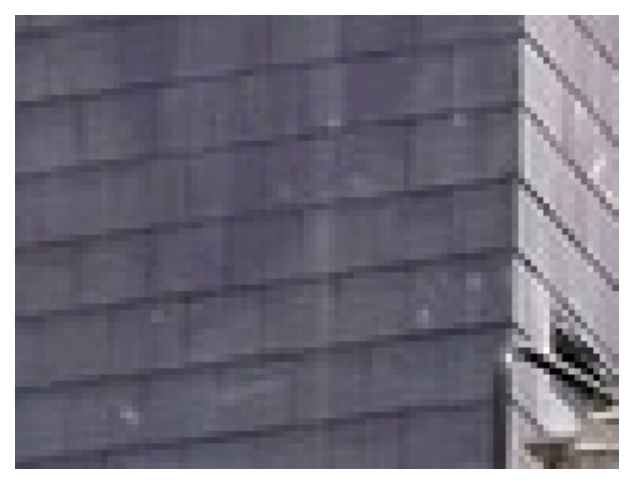}
        \caption*{HR (img-53)} 
    \end{subfigure}
\end{subfigure}
\begin{subfigure}{0.155\textwidth} 
    \begin{subfigure}{\textwidth}
        \includegraphics[width=\textwidth]{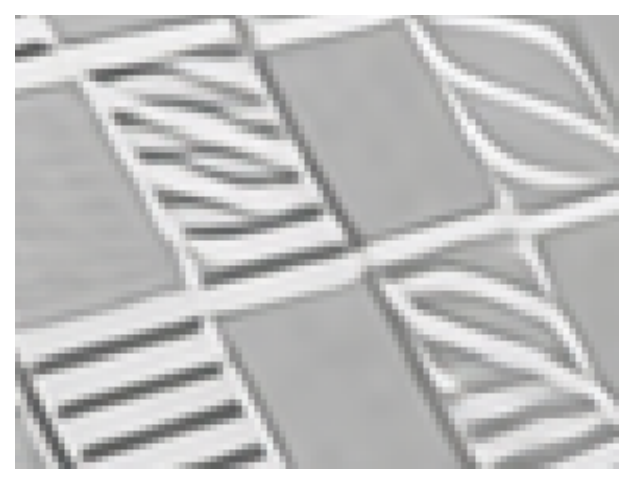}
        \caption*{SwinIR \cite{Liang2021SwinIRIR}}
    \end{subfigure} 
\end{subfigure}
\begin{subfigure}{0.155\textwidth} 
    \begin{subfigure}{\textwidth}
        \includegraphics[width=\textwidth]{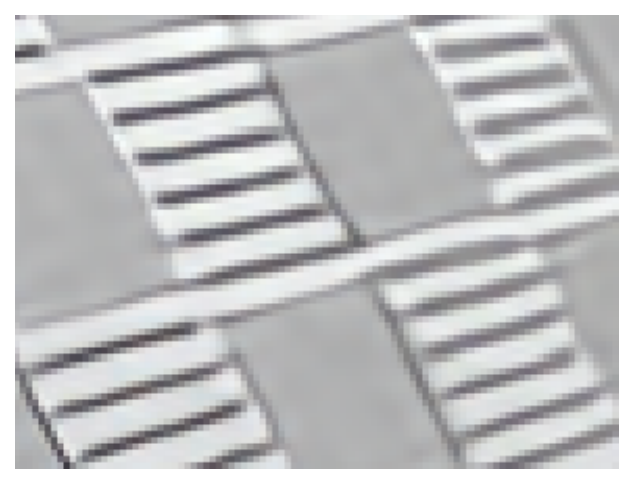}
        \caption*{SwinIR + SWT}
    \end{subfigure} 
\end{subfigure}
\begin{subfigure}{0.155\textwidth} 
    \begin{subfigure}{\textwidth}
        \includegraphics[width=\textwidth]{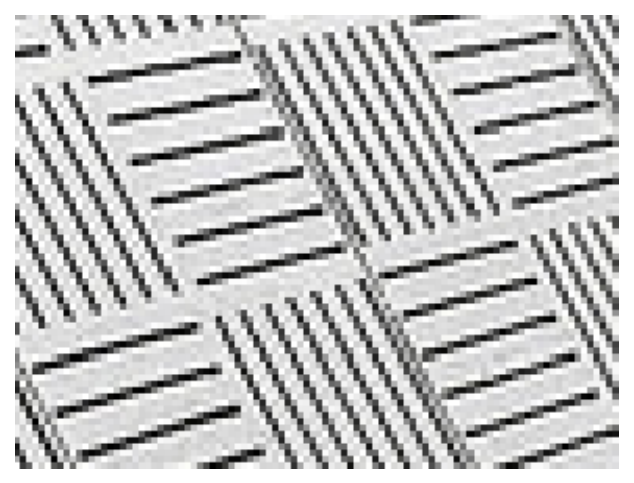}
        \caption*{HR (img-92)}
    \end{subfigure} 
\end{subfigure} \vspace{-18pt}
\caption{Visual comparison of SwinIR trained by $l_1$ loss \cite{Liang2021SwinIRIR} vs. trained by SWT losses on images 53 \& 92 from Urban100 dataset \cite{urban100_cite}. Observe that training SwinIR by $l_1$ loss results in hallucinated edge directions, whereas SwinIR trained by weighted $l_1$ and SWT losses (SwinIR+SWT) recovers all structures correctly.}
\label{fig:wavelet_added} 
\end{figure}

\subsection{Power of Training by Wavelet Losses}
This subsection demonstrates that training by a weighted combination of RGB and SWT losses contributes to improved reconstruction quality with other Transformer-based SR models as well. We take SwinIR \cite{Liang2021SwinIRIR} as a case study in Figure~\ref{fig:wavelet_added}. Figure~\ref{fig:wavelet_added} shows SwinIR trained by $l_1$ loss only exhibits artifacts and erroneous reconstructions, notably in hallucinating roof bricks as parallel lines. However, incorporating the proposed wavelet losses in the training without altering any other configuration or additional training data, SwinIR+SWT results in more accurate reconstructions, particularly in preserving structural details akin to the ground-truth HR image. While training the SwinIR+SWT, we scale LL and HH subbands with 0.05, whereas LH and HL subbands are multiplied by 0.01. We observe that SwinIR+SWT showcases improved capabilities in overcoming aliasing artifacts and recovering correct line orientations in both images. 

{\it To summarize, training by wavelet losses improves the~performance of the SwinIR model \cite{Liang2021SwinIRIR} by mitigating artifacts without any bells and whistles.}

\begin{table*}[t!]
 \caption{Ablation studies to demonstrate the contributions of adding NLSA blocks and using wavelet loss in $\times$4 SR results on benchmarks.} \vspace{-3pt}
    \centering
    \scalebox{0.98}{
    \begin{tabular}{llllll}
    \specialrule{.1em}{.05em}{.05em} 
    & \hspace{18pt} Baseline & \hspace{20pt} Baseline & \hspace{12pt} +NLSA blocks & \hspace{20pt} +$L_{SWT}$ & \hspace{12pt} +NLSA+ $L_{SWT}$ \\  
    \hline
    & \hspace{18pt} ImageNet & \hspace{20pt} ImageNet & \hspace{12pt} ImageNet & \hspace{20pt} ImageNet& \hspace{12pt} ImageNet \\  
    Training Set & \hspace{18pt} +DF2K & \hspace{20pt} +DF2K & \hspace{12pt} +DF2K & \hspace{20pt} +DF2K & \hspace{12pt} +DF2K\\ 
    &  & \hspace{20pt} +LSDIR & \hspace{12pt} +LSDIR & \hspace{20pt} +LSDIR & \hspace{12pt} +LSDIR\\ 
    \hline
    Benchmark & \hspace{10pt} PSNR - SSIM & \hspace{12pt} PSNR - SSIM & \hspace{14pt} PSNR - SSIM & \hspace{14pt} PSNR - SSIM & \hspace{16pt} PSNR - SSIM \\ 
    \hline
    Set5 & \hspace{6pt} 33.039 - 0.9054 & \hspace{12pt} 33.143 - 0.9071 & \hspace{12pt} 33.204 - 0.8893 & \hspace{12pt} 33.268 - 0.9084 &\hspace{16pt} 33.273 - 0.9082\\
    Set14 & \hspace{6pt} 29.246 - 0.7974 & \hspace{12pt} 29.387 - 0.8007 & \hspace{12pt} 29.524 - 0.8071 & \hspace{12pt} 29.527 - 0.8020  &\hspace{16pt} 29.524 - 0.8020 \\
    BSD100 & \hspace{6pt} 27.996 - 0.7515 & \hspace{12pt} 28.091 - 0.7547 & \hspace{12pt} 27.795 - 0.7422 & \hspace{12pt} 28.118 - 0.7548 &\hspace{16pt} 28.119 - 0.7549 \\
    Urban100 & \hspace{6pt} 27.954 - 0.8366 & \hspace{12pt} 28.247 - 0.8299 & \hspace{12pt} 28.548 - 0.8434 & \hspace{12pt} 28.684 - 0.8504  &\hspace{16pt} 28.692 - 0.8506 \\
    DIV2K & \hspace{6pt} 31.219 - 0.8547 & \hspace{12pt}  31.393 - 0.8575 & \hspace{12pt} 31.363 - 0.8482 & \hspace{12pt} 31.426 - 0.8577 &\hspace{16pt} 31.430 - 0.8578\\
    LSDIR & \hspace{6pt} 26.995 - 0.7883 & \hspace{12pt} 27.247 - 0.7955 & \hspace{12pt} 26.879 - 0.7780& \hspace{12pt} 27.292 - 0.7956 &\hspace{16pt} 27.296 - 0.7958\\
\specialrule{.1em}{.05em}{.05em} 
    \end{tabular} 
}
\label{table:ablation_results}
\end{table*}

\begin{table}[h!]
 \caption{Analysis of the effect of the number of SWT-decomposition levels used to compute the wavelet loss term on the~model performance over the BSD100 \cite{bsd100_cite} dataset.} 
    \centering
    \scalebox{1}{
    \begin{tabular}{lll}
    \specialrule{.1em}{.05em}{.05em} 
    Benchmark & \hspace{0.5cm} BSD100 & \hspace{0.3cm} Urban100\\ \hline
    Method & PSNR - SSIM & PSNR - SSIM \\ \hline
    Baseline & 28.088 - 0.756 & 28.589 - 0.849 \\
    1-level SWT & 28.119 - 0.755 & 28.697 - 0.851 \\
    2-level SWT & 27.514 - 0.742 & 28.699 - 0.851\\
    \specialrule{.1em}{.05em}{.05em} 
    \end{tabular} 
}
\label{table:wv_level}
\end{table}

\subsection{The Effect of Wavelet-Decomposition Levels}
The selection of the number of levels in the SWT decomposition plays a crucial role and directly affects the overall performance of the SR outputs. This decision is influenced by various factors, including the scale and orientation of structures within LR images. Our investigation focuses on assessing the impact of the decomposition level within our auxiliary wavelet-loss term. Specifically, we experiment with a 2-level SWT loss by further decomposing the LL subband of the 1-level SWT into 4 subbands (L-LL, L-LH, L-HL, L-HH), resulting in a total loss term that involves 8 distinct pixel-wise loss calculations (1 for RGB pixel-loss and 7 for wavelet subbands). Our findings, detailed in Table \ref{table:wv_level}, indicate that while the utilization of a 2-level wavelet loss does not significantly improve network generalization performance on the BSD100 \cite{bsd100_cite} benchmark, it does lead to performance improvements on the Urban100 \cite{urban100_cite} dataset. 

{\it To summarize, the effect of the number of SWT decomposition levels varies depending on the properties of images, such as the scale/frequency and orientation of structures, in the training/test datasets.}

\subsection{Ablation Study}
In this subsection, we conduct a series of ablation study experiments to analyze the individual impact of fine-tuning the HAT-L architecture with each of the following steps: i)~using the large-scale LSDIR dataset \cite{LSDIR_Li_2023_CVPR}, ii) inclusion of NLSA blocks, and iii) our training strategy with wavelet losses. The evaluation is performed on six benchmark datasets, including Set5 \cite{set5_cite}, Set14 \cite{set14_cite}, BSD100 \cite{bsd100_cite}, Urban100 \cite{urban100_cite}, DIV2K \cite{Agustsson_2017_CVPR_Workshops}, and LSDIR \cite{LSDIR_Li_2023_CVPR}, with results presented in Table \ref{table:ablation_results}. 

We first fine-tune our baseline model HAT-L \cite{chen2023activating} with the recently published large-scale dataset LSDIR containing diverse natural images \cite{LSDIR_Li_2023_CVPR}. This fine-tuning step yields a gain of +0.1 dB across all benchmarks, prompting us to utilize LSDIR dataset for the other experiments, including sandwiching in between the NLSA blocks and training with wavelet losses. Hence, as a baseline model, we consider HAT-L architecture that is fine-tuned on LSDIR \cite{LSDIR_Li_2023_CVPR} dataset. Subsequently, we introduce +4 NLSA blocks \cite{mei_nlsa} to the baseline architecture to augment the receptive field and the inclusion of NLSA blocks improves Set5 \cite{set5_cite}, Set14 \cite{set14_cite} and Urban100 \cite{urban100_cite} datasets up to 0.3 dB. Then, we train baseline architecture with SWT loss only without including NLSA blocks which significantly enhances overall results across all benchmarks, with an improvement of approximately 0.2 dB. Additionally, by incorporating the NLSA block and introducing an auxiliary wavelet loss function to refine our Wavelettention model, we achieve a notable enhancement, demonstrating a significant PSNR gain of 0.32 dB across almost every validation benchmark, surpassing the baseline performance.

\subsection{NTIRE 2024 Single Image Super-Resolution (x4) Challenge Results}
With the proposed Wavelettention model, we participated in the Image Super-Resolution $\times$4 subtrack of the New Trends in Image Restoration and Enhancement (NTIRE)~2024 challenge \cite{chen2024ntire_sr}. This challenge aims to design SR methods with the best PSNR and SSIM performance. It comprises of two datasets: DIV2K \cite{Agustsson_2017_CVPR_Workshops} and LSDIR \cite{LSDIR_Li_2023_CVPR}. Specifically, the~DIV2K training dataset  contains 800 pairs of high-resolution (HR) and low-resolution (LR) images and the~validation set consists of 100 LR-HR pairs. During the~final phase of the challenge, DIV2K test dataset containing 100 diverse LR images has been released to generate SR results. Our Wavelettention SR model actively participated in both the validation and testing phases of this challenge with the outcomes shown in Table \ref{table:ntire_results}.

\begin{table}
 \caption{NTIRE 2024 Image Super-Resolution Challenge Results.  PSNR and SSIM scores for $\times$4 SR on validation and test phases.} \vspace{-6pt}
    \centering
    \scalebox{1}{
    \begin{tabular}{lcc}
    \specialrule{.1em}{.05em}{.05em} 
    & PSNR & SSIM \\  
    \hline
    Validation Dataset & 31.43 & 0.86 \\
    Test Dataset  & 31.13 & 0.86 \\
\specialrule{.1em}{.05em}{.05em} 
    \end{tabular} 
}
\label{table:ntire_results}
\end{table}

\subsection{Analysis of Model Complexity vs. Performance}
We conduct experiments to analyze the computational complexity versus performance gain of additional NLSA blocks to the baseline HAT-L \cite{chen2023activating} architecture. As illustrated in Table \ref{table:nlsa_block_num}, adding single NLSA block before and after the feature extraction layers of HAT-L \cite{chen2023activating} model increases model performance 0.14 dB on Set14 \cite{set14_cite}. Furthermore, our Wavelettention model that contains +4 NLSA blocks obtains a performance gain by 0.2 dB with an increase of parameters and Multi-Adds. However, further addition of NLSA blocks does not lead to a performance gain even though the parameter size and Multi-Add increases. This problem may be addressed by providing additional training data. 

{\it To summarize, we observe that inserting 2 or 4 NLSA blocks both before and after the feature extraction module of HAT-L leads to an increase in the performance of the state-of-the-art HAT model 
with a small increase in complexity.}

\begin{table}[h!]
 \caption{Complexity vs. performance comparison for $\times$4 SR. Each row shows the cost and benefit of adding NLSA blocks \cite{mei_nlsa} to the baseline HAT-L \cite{chen2023activating} architecture on Set14 \cite{set14_cite} benchmark.} 
    \centering
    \scalebox{0.9}{
    \begin{tabular}{lccc}
    \specialrule{.1em}{.05em}{.05em} 
    Method & $\#$ Params. & $\#$ Multi-Adds. & PSNR \\
    \hline
    Baseline & 40.8M & 79.61G & 29.349 \\
    +2 NLSA Blocks & 41.1M & 80.47G & 29.485 \\
    +4 NLSA Blocks & 41.3M & 81.33G & 29.524 \\
    +8 NLSA  Blocks & 41.7M & 83.06G & 29.496 \\
\specialrule{.1em}{.05em}{.05em} 
    \end{tabular} 
}
\label{table:nlsa_block_num}
\end{table}

\section{Conclusion}

This paper introduces a new hybrid transformer model for image SR by sandwiching a transformer architecture in between convolutional NLSA blocks aiming to further increase the receptive field of the model and enhance the quality of the generated SR images. Furthermore, we address the well-known limitation of RGB pixel-wise losses in capturing high-frequency details that are crucial for visual quality, by introducing wavelet subband losses. Specifically, we propose training SR models by a weighted combination of RGB and wavelet losses to preserve the scale and orientation of high-frequency image details for improved SR performance. Through extensive ablation studies, we show that both the proposed architectural improvement and use of wavelet losses in training help us achieve better quantitative~(PSNR and SSIM) scores and qualitative (visual) results compared to state-of-the-art transformer-based methods for the SR task. We also demonstrated that training by wavelet losses can improve the performance of other transformer-based SR models, such as the SwinIR model.

\section{Acknowledgements} 
This work is supported by TUBITAK 2247-A Award No. 120C156, TUBITAK 2232 International Fellowship for Outstanding Researchers Award No. 118C337, KUIS AI Center, and Turkish Academy of Sciences (TUBA).

{
    \small
    \bibliographystyle{ieeenat_fullname}
    \bibliography{source}
}


\end{document}